\documentclass[10pt,twocolumn,letterpaper]{article}

\usepackage[pagenumbers]{cvpr}

\definecolor{cvprblue}{rgb}{0.21,0.49,0.74}
\usepackage[pagebackref,breaklinks,colorlinks,allcolors=cvprblue]{hyperref}

\title{Motion-Robust Deep Reconstruction for Free-Breathing Cardiac Cine MRI}

\author{Mahmut Yurt\\
Stanford University\\
\and
Kanghyun Ryu\\
KIST\\
\and
Zhitao Li\\
Stanford University\\
\and
Xucheng Zhu\\
GE Healthcare\\
\and
Xianglun Mao\\
GE Healthcare\\
\and
Martin Janich\\
GE Healthcare\\
\and
Marcus Alley\\
Stanford University\\
\and
Kawin Setsompop\\
Stanford University\\
\and
John Pauly\\
Stanford University\\
\and
Shreyas Vasanawala\\
Stanford University\\
\and
Ali Syed\\
Stanford University\\
}

\begin{document}
\maketitle
\begin{abstract}
Conventional cardiac cine MRI relies on breath-hold Cartesian acquisitions, which are vulnerable to motion artifacts and can be uncomfortable or infeasible, particularly for pediatric and other noncompliant patients who cannot reliably hold their breath. Free-breathing radial acquisitions can alleviate these limitations, but robust reconstruction at high acceleration remains challenging due to prominent streak artifacts. To address these limitations, we propose Cine-DL, a clinically oriented framework that couples targeted k-space preprocessing with fast, model-based deep reconstruction. In this pipeline, raw free-breathing radial data undergo retrospective cardiac binning and respiratory gating to resolve cardiac phases and discard motion-corrupted spokes. We then introduce Streak Optimized Coil Compression (SOC), which explicitly preserves cardiac signals while suppressing peripheral interference that typically drives the streak artifacts. The resulting 2D+t cine series is reconstructed with an unrolled network that alternates a ResNet proximal operator with physics-based data consistency updates solved via conjugate gradient. We further employ a memory-efficient training strategy that reduces peak memory usage. We evaluate Cine-DL on free-breathing volunteer data against established baselines (\textit{k-t} SENSE and iGRASP) and demonstrate clinical translation via hospital deployment on newly acquired patient data. Our experiments show that Cine-DL consistently improves quantitative metrics and visual fidelity, supporting a practical route toward routine, time-sensitive clinical adoption of free-breathing cine MRI.
\end{abstract}
    
\section{Introduction}
\label{sec:intro}

Cardiac magnetic resonance imaging (CMR) is a central modality for non-invasive assessment of cardiac morphology and function, offering excellent soft-tissue contrast and the ability to prescribe images in arbitrary, patient-specific planes \cite{ismail2022cardiac}. Cine CMR further provides dynamic imaging of the beating heart as a $2\text{D}+t$ sequence spanning the cardiac cycle. These capabilities make cine CMR a routine clinical tool for diagnosis and longitudinal follow-up across a broad range of cardiovascular diseases \cite{sechtem1987cine, authors2012esc}.

However, conventional cine CMR is typically acquired with Cartesian k-space sampling, which is inherently sensitive to motion-induced inconsistencies \cite{zaitsev2015motion, feng2022golden}. In the presence of respiratory motion, these inconsistencies appear as blurring and ghosting, motivating widespread use of breath-hold acquisitions, where data are collected over multiple short breath-holds (often across several heartbeats) to reduce respiratory motion. Breath-holding, in turn, can reduce patient comfort and is often infeasible in pediatric, elderly, or dyspneic patients, limiting scan duration and thereby constraining achievable image quality \cite{vasanawala2010improved}.

Free-breathing golden-angle radial sampling is a promising alternative to conventional Cartesian cine acquisitions, enabling continuous acquisition of radial spokes without breath-holds and thereby improving patient comfort and compliance \cite{feng2022golden, leung2008free, kravchenko2023free}. By repeatedly sampling the center of the k-space and rotating spoke directions over time, radial acquisitions distribute motion-induced inconsistencies more incoherently, improving tolerance to motion and reducing structured artifacts \cite{zaitsev2015motion}. In addition, golden-angle ordering supports flexible and balanced retrospective binning of the spokes, which is well suited to reconstruction of dynamic cine imaging.

Despite these advantages, high-quality radial cine reconstruction remains challenging. Radial undersampling can produce streak artifacts, frequently driven by high-intensity peripheral signals and exacerbated by system imperfections \cite{xue2012automatic}. Moreover, accurate reconstruction typically depends on NUFFT-based non-Cartesian forward models and iterative optimization, which can be computationally intensive and may yield limited reconstruction quality under high acceleration, reducing practicality in time-sensitive and accuracy-critical workflows in the clinic.

In this work, we present a clinically oriented acquisition and reconstruction pipeline for free-breathing cine CMR that couples targeted k-space processing with a fast, model-based deep reconstruction network. Our framework (i) uses golden-angle radial sampling to enable continuous free-breathing acquisition, (ii) suppresses radial streak artifacts via Streak Optimized Coil Compression (SOC) that prioritizes the central cardiac region while attenuating peripheral interference, and (iii) reconstructs $2\text{D}+t$ cardiac-phase cine series using an unrolled network that alternates a learned ResNet proximal operator with physics-based data-consistency updates. To make training practical for this high-dimensional reconstruction task, we adopt a memory-efficient unrolled training strategy that substantially reduces GPU memory usage while preserving end-to-end learning. 

We evaluate the approach in two stages: first on volunteer data against standard reconstruction baselines, and then through clinical deployment, where we assess the same pipeline on newly acquired clinical patient data from a local clinic. Across both short-axis stacks and four-chamber views, our experiments demonstrate consistent improvements over baselines in both quantitative metrics and qualitative image quality, supporting the clinical potential of the proposed free-breathing radial cine framework.
\\~\\
\textbf{Contributions. }Our contributions are as follows:
\\
\begin{itemize}
    \item We propose a clinically-oriented acquisition and reconstruction framework for free-breathing radial cine CMR.
    \item We introduce Streak Optimized Coil Compression (SOC) to reduce streak artifacts and emphasize cardiac regions.
    \item We propose a fast unrolled reconstruction with a ResNet proximal and physics-based data consistency, trained with memory efficient learning.
    \item We performed extensive validation on volunteer and patient data and deployment in clinical settings.
\end{itemize}
\section{Related Work}
\label{sec:related_work}

\textbf{Accelerated dynamic MRI:} Earlier work on dynamic MRI reconstruction explored redundancy across space and time. \textit{k-t} BLAST and \textit{k-t} SENSE \cite{tsao2003k} model the signal distribution in the spatiotemporal frequency domain and recover missing measurements by learning the correlations, enabling acceleration for quasi-periodic motion such as cardiac imaging. Meanwhile, iGRASP \cite{feng2014golden} combines parallel imaging with compressed sensing and leverages a temporal total-variation (TV) constraint to suppress residual aliasing and streaking artifacts. While iGRASP improves image quality, its temporal TV regularization can suppress small temporal coefficients and potentially impact temporal fidelity at aggressive accelerations.
\\~\\
\textbf{Coil processing for artifact reduction: }Region-optimized virtual (ROVir) coils form linear combinations of channels to suppress signals from predefined interference regions while preserving signal in the region of interest \cite{kim2021region, haldar2023optimality}. The mixing weights are obtained by maximizing a signal-to-interference ratio (SIR) via a generalized eigendecomposition. Radial acquisitions are especially prone to streaks from strong peripheral signal and coils with high sensitivity near the field-of-view edges. On the other hand, coil removal \cite{xue2012automatic} is an automated coil-selection method that detects coils with severe streaking and excludes them from reconstruction using a streaking artifact ratio metric, formed by comparing a full-resolution coil image to a streak-suppressed low-resolution reference obtained via low-pass Hanning filtering in k-space and thresholding this metric. 
\\~\\
\textbf{Model-based deep reconstruction and unrolled networks: }Model-based deep learning unrolls iterative reconstruction into a network that alternates learned regularization with explicit data consistency steps. MoDL \cite{aggarwal2018modl} is an important method that combines a CNN prior with a physics-based data consistency layer. For multi-coil models, it typically uses conjugate gradient (CG) and often shares weights across iterations to reduce parameters. Related unrolled approaches also include DL-ESPIRiT \cite{sandino2021accelerating}, which uses an ESPIRiT-style multi-map sensitivity model in data consistency to handle reduced-FOV acquisitions \cite{uecker2014espirit}, together with separable 2D spatial and 1D temporal convolutions \cite{sandino2021accelerating}. For high-dimensional unrolled reconstructions, memory can limit the depth of the unrolls. A recent study \cite{wang2021memory} introduced memory-efficient learning (MEL), which reduces storage by recomputing intermediate variables during backpropagation and storing only one block at a time, enabling deeper networks and improved fidelity. More recently, another study proposed GLEAM \cite{ozturkler2022gleam}, a greedy training strategy that splits the end-to-end unrolled network into decoupled modules optimized with local losses, where intermediate activations are discarded between modules to reduce training memory, and the decoupled updates can be parallelized across multiple GPUs \cite{ozturkler2022gleam}.
\section{Methods}
\label{sec:methods}

\begin{figure*}[htb]
    \centering
    \includegraphics[width=0.99\textwidth]{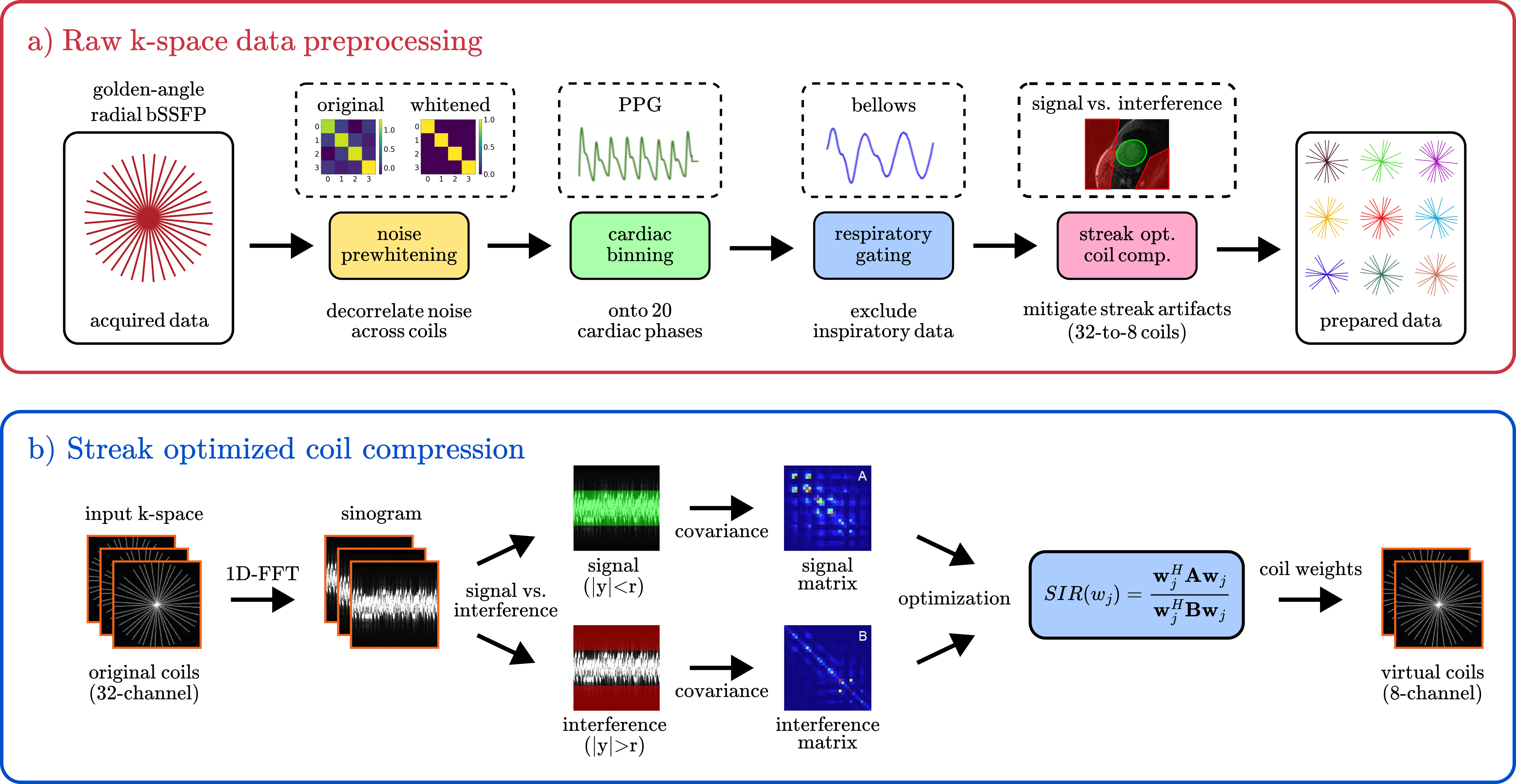}
    \caption{\textbf{Overview of the proposed k-space preprocessing and streak optimized coil compression (SOC).}
a) Raw golden-angle radial data are first noise-prewhitened to decorrelate coil noise, then retrospectively cardiac-binned using PPG/ECG into $T{=}20$ phases and respiratory-gated using the bellows signal to reject inspiratory spokes. The resulting prepared multi-coil, phase-binned data are passed to coil compression to reduce streak artifacts and computational load.
b) SOC pipeline: multi-coil k-space is transformed to the sinogram domain via a 1D FFT along readout, partitioned into signal and interference regions, and used to form corresponding covariance matrices. Coil combination weights are obtained by maximizing a signal-to-interference ratio (SIR) criterion, yielding a reduced set of virtual coils.}
\label{fig:preproc_roc}
\end{figure*}

\subsection{Raw k-space preprocessing}
\label{sec:raw_kspace_processing}

\textbf{Input representation: }Raw multi-coil k-space data are represented as $y \in \mathbb{R}^{N_{RO}\times N_{sp}\times N_c \times N_{sl}}$, where $N_{RO}$ is the readout length of the spokes, $N_{sp}$ the number of spokes per slice, $N_c$ the number of coils, and $N_{sl}$ the number of slices. Prior to downstream processing, we perform coil noise prewhitening to decorrelate noise across coil channels. For each slice, we also generate the corresponding 2D non-Cartesian radial trajectory $\Omega \in \mathbb{R}^{N_{sp}\times N_{RO}\times 2}$ with coordinates in $(k_y,k_x)$ ordering to match the operator convention by our reconstruction.
\\~\\
\textbf{Cardiac binning and respiratory gating: }Physiologic traces (ECG/PPG triggers for cardiac cycle and respiratory bellows for respiration) are used to retrospectively assign each spoke to a cardiac phase and to reject spokes corrupted by large respiratory motion. Specifically, for each slice, we partition spokes into $T=20$ cardiac bins and represent the binning as an index set $\{\mathcal{I}_{t}\}_{t=1}^{T}$, where $\mathcal{I}_{t}\subset\{1,\dots,N_{sp}\}$ denotes the spokes assigned to phase $t$. This yields per-bin data $y_t \in \mathbb{R}^{N_{RO}\times N_{sp}^{(t)}\times N_c}$
with $N_{sp}^{(t)} = |\mathcal{I}_t|$ spokes per phase. Respiratory gating is applied using the bellows signal by excluding spokes whose respiratory surrogate exceeds a fixed threshold (corresponding bin indices are filtered by the gating mask) prior to reconstruction.

\subsection{Streak Optimized Coil Compression (SOC)}
\label{sec:coilcompression}

To reduce computational cost while mitigating radial streaking, we apply Streak Optimized Coil Compression (SOC) to project the original $N_c$ coils onto a smaller set of $N_v$ virtual coils. Unlike standard SVD coil compression, which primarily maximizes SNR over the full field-of-view and may result in strong peripheral signal \cite{zhang2013coil}, SOC explicitly prioritizes the preservation of the cardiac regions and suppresses peripheral non-cardiac contributions that commonly drive streak artifacts in radial cine.
\\~\\
\textbf{Sinogram-based region definition: }For each slice, we map the acquired radial k-space to the projection (sinogram) domain by applying a 1D Fourier transform along the readout dimension, yielding coil-resolved sinograms. We then define two complementary sets of sinogram samples: (i) a signal region that captures the central cardiac content, and (ii) an interference region dominated by peripheral anatomy that commonly generates streaking.
\\~\\
\textbf{SIR-optimized compression: }Let $\{\mathbf{s}_n\}_{n\in\mathcal{S}}$ and $\{\mathbf{i}_m\}_{m\in\mathcal{I}}$ denote the $N_c$-dimensional coil sample vectors drawn from the signal and interference regions, respectively. We estimate the corresponding covariance matrices $\mathbf{A}=\frac{1}{|\mathcal{S}|}\sum_{n\in\mathcal{S}}\mathbf{s}_n\mathbf{s}_n^{H}$
and $\mathbf{B}=\frac{1}{|\mathcal{I}|}\sum_{m\in\mathcal{I}}\mathbf{i}_m\mathbf{i}_m^{H}$,
and compute coil combination vectors by maximizing $\mathrm{SIR}(\mathbf{w})=\frac{\mathbf{w}^{H}\mathbf{A}\mathbf{w}}{\mathbf{w}^{H}\mathbf{B}\mathbf{w}}$.
This yields virtual-coil weights that preserve cardiac signal while suppressing peripheral interference.

\subsection{Unrolled network reconstruction}
\label{sec:unrolled}

\begin{figure*}[htb]
    \centering
    \includegraphics[width=0.83\textwidth]{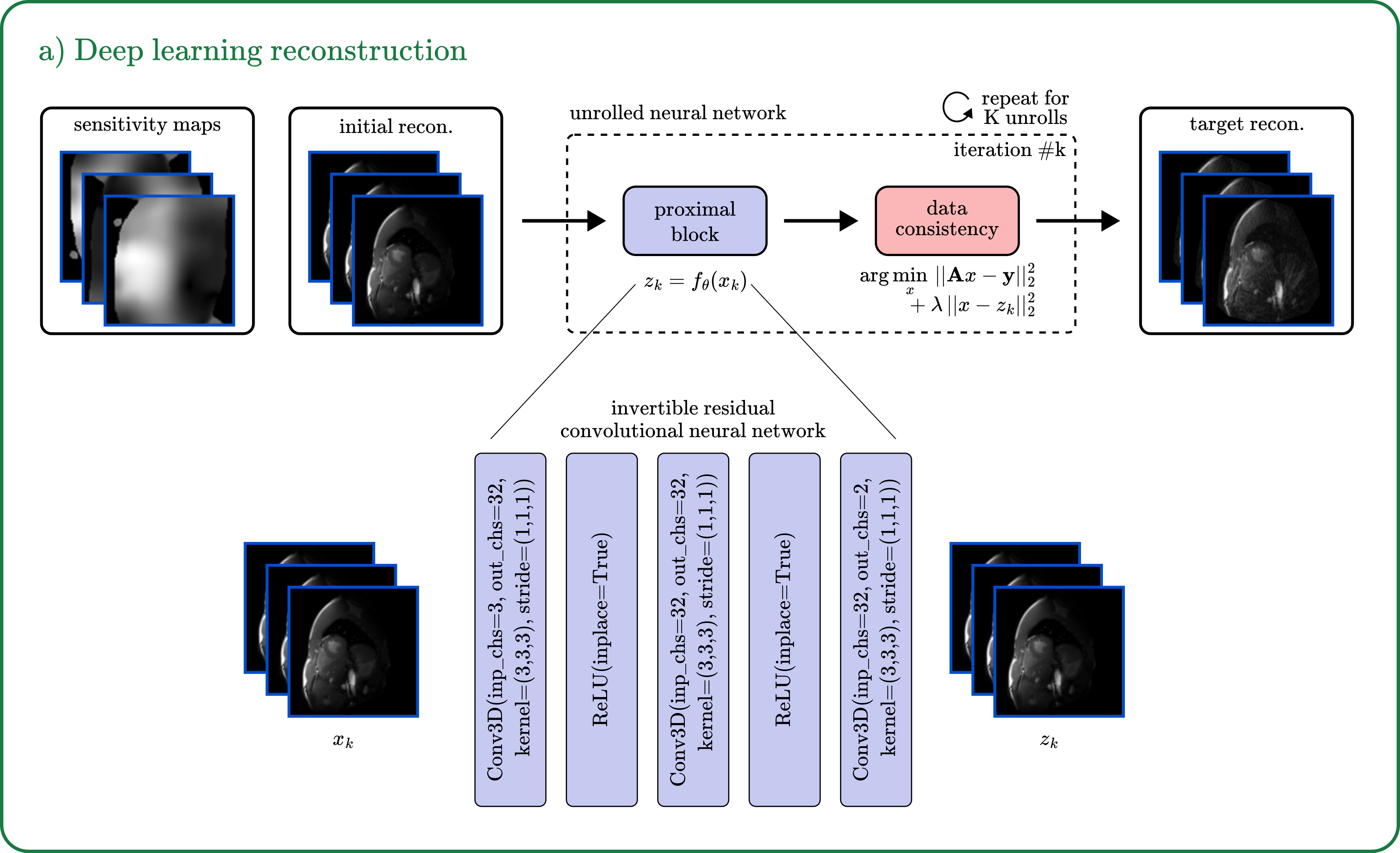}
    \caption{\textbf{Deep learning reconstruction with an unrolled network.}
Given coil sensitivity maps and an initial reconstruction, we apply an unrolled neural network for $K$ iterations. Each unroll alternates a learned proximal block with a physics-based data-consistency update that enforces agreement with the acquired non-Cartesian k-space data. The proximal block refines the current estimate $x_k$ to produce an intermediate denoised/artifact-suppressed image $z_k$, after which the data-consistency step produces the next iterate $x_{k+1}$.}
\label{fig:unrolled_dl}
\end{figure*}

\textbf{Forward model and sensitivity maps: }
For each cardiac phase $t\in\{1,\dots,T\}$, we reconstruct a complex image $x_t \in \mathbb{R}^{N\times N}$ from the binned non-Cartesian measurements $y_t$ using a multi-coil SENSE forward model \cite{pruessmann1999sense}:
\begin{equation}
y_t = A_t x_t + \eta,
\end{equation}
where $A_t$ composes coil sensitivity encoding, NUFFT \cite{fessler2007nufft} sampling on the bin-specific radial trajectory, and phase-specific spoke selection with DCF weighting. Coil sensitivity maps are estimated using JSENSE \cite{ying2007joint}. Before JSENSE, we apply spoke-wise phase correction and pre-weight k-space by DCF computed on the selected spokes. For deep inference, maps are additionally normalized to have unit root-sum-of-squares magnitude across coils.
\\~\\
\textbf{Unrolled reconstruction with learned proximal and data consistency: }
As illustrated in Fig.~\ref{fig:unrolled_dl}, the unrolled network maintains an iterate $x_k$ (the current reconstruction estimate) and applies a learned proximal block to produce
\begin{equation}
z_k = f_{\theta}(x_k),
\end{equation}
followed by a physics-based data-consistency update that computes the next iterate $x_{k+1}$ by approximately solving
\begin{equation}
x_{k+1} = \arg\min_{x}\ \|A x - y\|_2^2 + \lambda \|x - z_k\|_2^2,
\end{equation}
using a small fixed number of conjugate-gradient (CG) iterations. This operator-based update enforces fidelity to the acquired k-space without explicit matrix formation. The proximal and data-consistency steps are repeated for $K$ unrolls.
\\~\\
\textbf{Learned proximal network (ResNet): }
The proximal $f_{\theta}(\cdot)$ is a lightweight 3D ResNet \cite{he2016deep} operating on the complex cine sequence represented as two channels (real and imaginary) and processing the full cardiac-phase dimension jointly. We use circular padding in the $(t,x,y)$ dimensions and append a scalar time-embedding channel $k/K$ to the input of each unroll to allow iteration-dependent behavior. The ResNet outputs a residual update that is added to its input, encouraging progressive artifact suppression while preserving fine cardiac structure \cite{he2016deep}.
\\~\\
\textbf{Memory-efficient training: }Non-Cartesian reconstruction relies on NUFFT operators and physics-based data-consistency steps that are significantly more memory demanding than Cartesian counterparts, which can constrain both network depth and model capacity in unrolled networks. To enable deeper unrolling without prohibitive memory usage, we employ a memory-efficient learning (MEL) framework \cite{wang2021memory} that trades additional computation for substantially reduced activation storage. Concretely, during training we store only the intermediate inputs and outputs of each unrolled block rather than the full computational graph, and reconstruct required intermediate states during backpropagation. This strategy reduces peak GPU memory while preserving end-to-end training of the unrolled reconstruction network.

\subsection{Datasets}
\label{sec:datasets}

\textbf{Volunteer dataset: }Free-breathing golden-angle radial cine short-axis stacks were collected from 11 volunteers with IRB approval and written informed consent, with 8 subjects for training, 1 for validation, and 2 for testing. From each subject, 11-15 short-axis slices were acquired with tiny golden angle using a slice thickness of 6mm, and 70-80 seconds of data per slice. Four-chamber views were additionally acquired from 1 subject for testing. Respiratory bellows and PPG signals were recorded for respiratory gating (to exclude inspiratory data) and cardiac binning (20 phases).
\\~\\
\textbf{Clinical dataset: }Free-breathing golden-angle radial cine data were acquired from 5 subjects at Lucile Packard Children’s Hospital at Stanford under IRB approval with written informed consent. For each subject, a variable number of short-axis slices were collected with acquisition durations ranging from 10 to 80 seconds per slice. Respiratory bellows and ECG signals were recorded for respiratory gating (excluding inspiratory data) and for cardiac binning into 20 phases.

\subsection{Metrics}
\label{sec:metrics}

We evaluate reconstruction quality using peak signal-to-noise ratio (PSNR), structural similarity index (SSIM), and streak-artifact ratio (SAR).
\\~\\
\textbf{PSNR: }Given a reconstruction $\hat{I}$ and reference image $I$, we compute the mean squared error
\begin{equation}
\mathrm{MSE} = \frac{1}{|\Omega|}\sum_{p\in\Omega}\left(I(p)-\hat{I}(p)\right)^2,
\end{equation}
and report PSNR (in dB) as
\begin{equation}
\mathrm{PSNR} = 20\log_{10}\!\left(\frac{\mathrm{MAX}_I}{\sqrt{\mathrm{MSE}}}\right),
\end{equation}
where $\mathrm{MAX}_I$ denotes the maximum intensity of the reference image.
\\~\\
\textbf{SSIM: }We report SSIM between $I$ and $\hat{I}$ as
\begin{equation}
\mathrm{SSIM}(I,\hat{I})=\frac{(2\mu_I\mu_{\hat{I}}+c_1)(2\sigma_{I\hat{I}}+c_2)}{(\mu_I^2+\mu_{\hat{I}}^2+c_1)(\sigma_I^2+\sigma_{\hat{I}}^2+c_2)},
\end{equation}
where $\mu$ and $\sigma^2$ denote local means and variances, $\sigma_{I\hat{I}}$ denotes local covariance, and $c_1,c_2$ are stabilization constants.
\\~\\
\textbf{Streak-artifact ratio (SAR): }To quantify streak contamination, we use the streak-artifact ratio \cite{xue2012automatic},
\begin{equation}
R_{\mathrm{SAR}} \triangleq 
\frac{\mathrm{mean}\!\left(\left|I_{\mathrm{org}}-I_{\mathrm{ref}}\right|\right)}
{\mathrm{mean}\!\left(I_{\mathrm{ref}}\right)},
\end{equation}
where $I_{\mathrm{org}}$ is the reconstructed image and $I_{\mathrm{ref}}$ is a ``streak-free'' low-resolution reference obtained by applying a low-pass Hanning filter in k-space.
\begin{figure}[htb]
    \centering
    \includegraphics[width=0.48\textwidth]{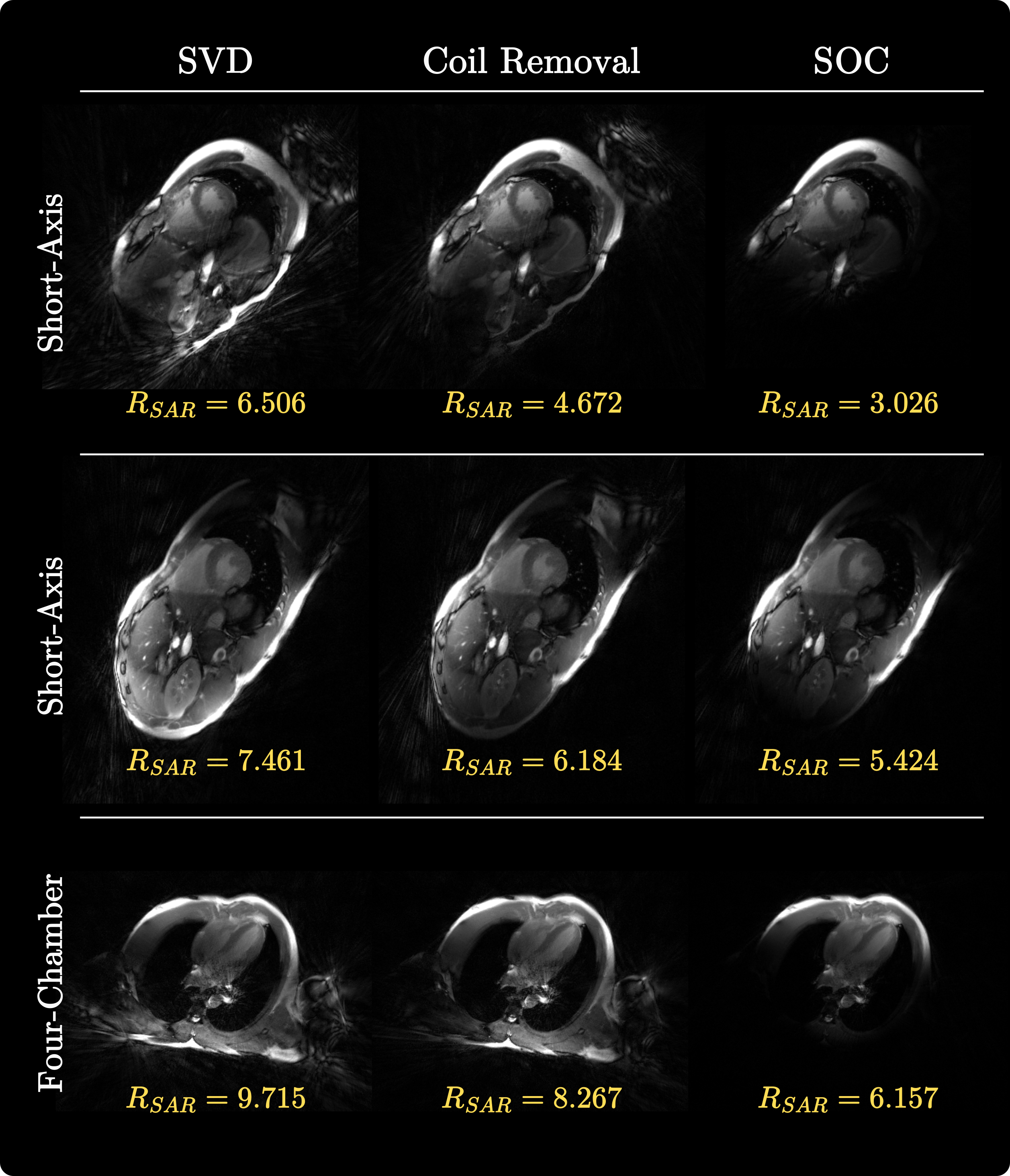}
\caption{\textbf{Efficacy of Streak Optimized Coil Compression (SOC).} Representative short-axis and four-chamber reconstructions after coil combination using (a) standard SVD coil compression, (b) coil removal~\cite{xue2012automatic}, and (c) the proposed SOC. By suppressing signal outside the heart, SOC reduces radial streak artifacts and limits their propagation into the myocardium and blood pool. The value reported beneath each image is the streak-artifact ratio ($R_{\text{SAR}}$), computed relative to a streak-suppressed low-resolution reference obtained by low-pass Hanning filtering in k-space (lower is better).}
\label{fig:roc_efficacy}
\end{figure}

\section{Experiments}
\label{sec:experiments}

\subsection{Efficacy of Streak Optimized Coil Compression}
\label{sec:roc_efficacy}

\textbf{Setup: }We first study how coil processing impacts streak artifacts and overall radial cine reconstruction quality. Keeping the acquisition, binning, and reconstruction settings fixed, we vary only the coil compression strategy and compare: \textbf{(i) standard SVD compression}, which projects the original multi-coil data onto a smaller set of virtual coils via conventional SVD; \textbf{(ii) coil removal}, which selects an automated subset of coils to reduce streaking using a streak metric from \cite{xue2012automatic}, followed by coil compression on the selected subset; and \textbf{(iii) proposed streak optimized coil compression (SOC)}, which explicitly preserves signal within the cardiac region while suppressing peripheral contributions that commonly drive radial streak artifacts via signal-to-interference ratio optimization. We evaluate performance across multiple slices in both short-axis and four-chamber views. We report qualitative comparisons using coil-combined images and quantify streak contamination per slice using the streak-artifact ratio $R_{\text{SAR}}$, computed against a streak-suppressed low-resolution reference obtained by low-pass Hanning filtering in k-space (please see Section \ref{sec:metrics} for details).
\\~\\
\textbf{Results: }
Representative results in Fig.~\ref{fig:roc_efficacy} show that the proposed SOC compression suppresses signal from non-cardiac regions more effectively than standard SVD compression and coil removal. By attenuating these peripheral contributions, which are the primary driver of radial streak artifacts, SOC produces visibly cleaner backgrounds. Importantly, it also reduces streaks that would otherwise propagate into the myocardium and blood pool. This qualitative improvement aligns with the quantitative trend across both short-axis and four-chamber views. In Fig.~\ref{fig:roc_efficacy}, $R_{\text{SAR}}$ decreases from $6.506 \rightarrow 3.026$, $7.461 \rightarrow 5.424$, and $9.715 \rightarrow 6.157$ relative to SVD, and also improves over coil removal ($4.672 \rightarrow 3.026$, $6.184 \rightarrow 5.424$, $8.267 \rightarrow 6.157$). Overall, SOC more effectively suppresses non-cardiac interference, yielding reconstructions with reduced streaking both globally and within the cardiac region of interest.

\begin{figure}[htb]
    \centering
    \includegraphics[width=0.48\textwidth]{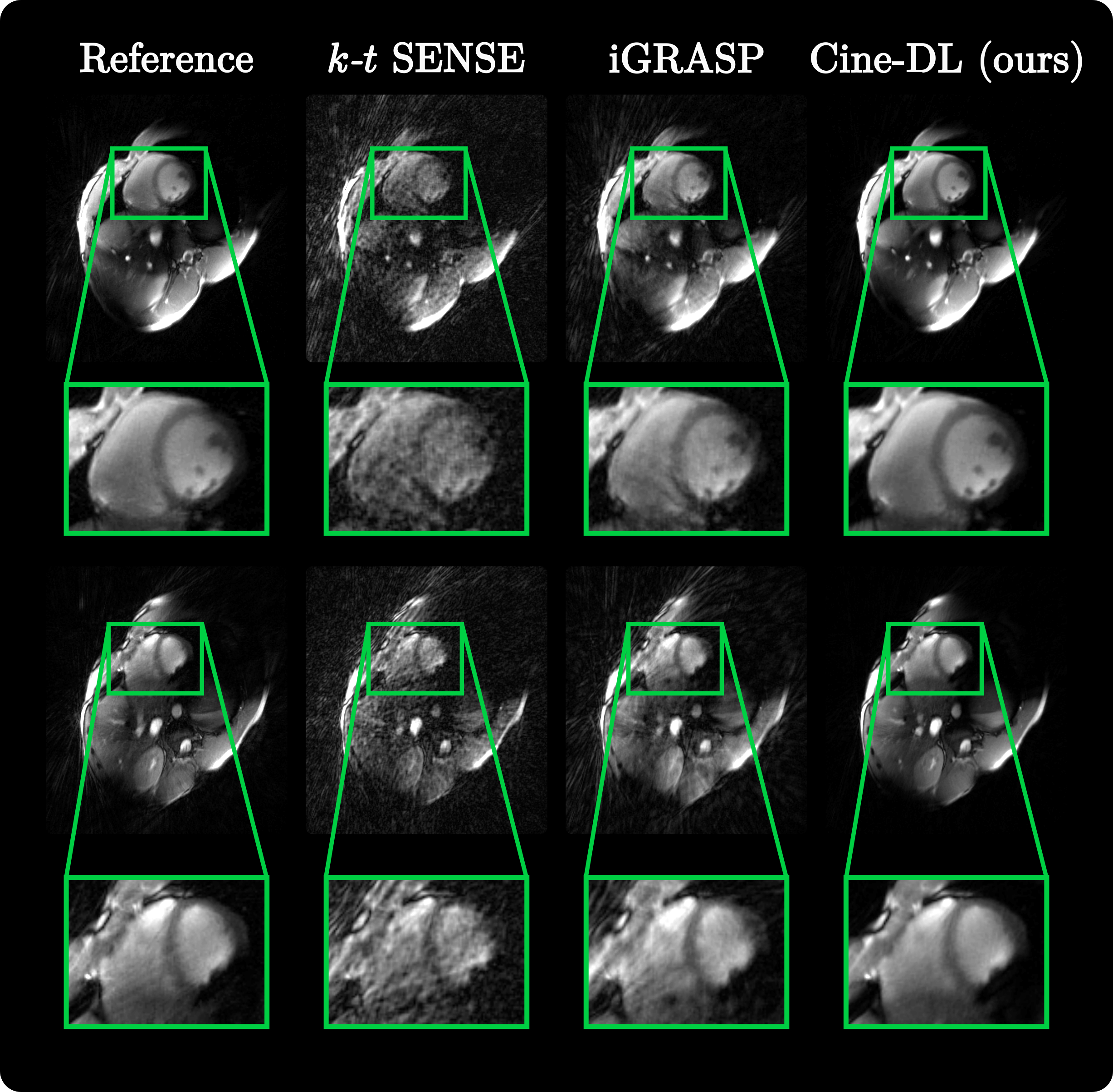}
\caption{\textbf{Visual comparison of free-breathing radial cine MRI reconstructions.} A high-quality reference acquisition (80~s/slice) is shown alongside reconstructions from \textit{k-t}~SENSE, iGRASP, and the proposed Cine-DL at an undersampling factor of R=6x (13~s/slice). Yellow boxes denote the cardiac regions with magnified insets illustrating that \textit{k-t}~SENSE exhibits residual aliasing/streaking and loss of fine detail, iGRASP suppresses artifacts but can introduce smoothing and residual artifacts, whereas Cine-DL yields cleaner images with sharper myocardial borders and fewer streaks propagating into the myocardium and blood pool.}
\label{fig:retrospective_results}
\end{figure}

\subsection{Demonstrations on volunteer data}
\label{sec:volunteer_results}

\begin{table*}[!t]
\centering
\renewcommand{\arraystretch}{1.5}
\resizebox{1.6\columnwidth}{!}{%
\begin{tabular}{l c@{\hspace{12pt}} c@{\hspace{24pt}} c@{\hspace{12pt}} c@{\hspace{24pt}} c@{\hspace{12pt}}c}
\hline\noalign{\hrule height 0.5pt}
         & \multicolumn{2}{c}{\textbf{R=4x}}  & \multicolumn{2}{c}{\textbf{R=6x}}  & \multicolumn{2}{c}{\textbf{R=8x}} \\ \cline{2-7}
         & \textbf{PSNR} & \textbf{SSIM} & \textbf{PSNR} & \textbf{SSIM} & \textbf{PSNR} & \textbf{SSIM} \\ \hline
\textit{k-t} SENSE & 31.92 $\pm$ 1.30 & 45.54 $\pm$ 6.46 & 31.10 $\pm$ 1.62 & 43.78 $\pm$ 6.81 & 31.11 $\pm$ 1.44 & 46.40 $\pm$ 6.66  \\ \hline
iGRASP   & 36.58 $\pm$ 1.97 & 82.71 $\pm$ 5.28 & 35.34 $\pm$ 2.36 & 77.23 $\pm$ 7.07 & 33.34 $\pm$ 2.36 & 70.22 $\pm$ 9.75  \\ \hline
Proposed & \textbf{39.15 $\pm$ 2.89} & \textbf{91.60 $\pm$ 4.60} & \textbf{38.76 $\pm$ 2.60} & \textbf{91.72 $\pm$ 3.65} & \textbf{38.48 $\pm$ 2.18} & \textbf{91.53 $\pm$ 3.55} \\ \hline\noalign{\hrule height 0.5pt}
\label{table:main}
\end{tabular}
}
\vspace{-5mm}
\caption{Quantitative measurements for the methods under comparison for radial cine MRI reconstruction under undersampling at acceleration factors R=4x, R=6x, and R=8x. We report PSNR (dB) and SSIM as mean $\pm$ standard deviation across test slices. Best results are shown in bold.}
\label{table:main}
\end{table*}

\begin{figure}[htb]
\centering
\includegraphics[width=0.48\textwidth]{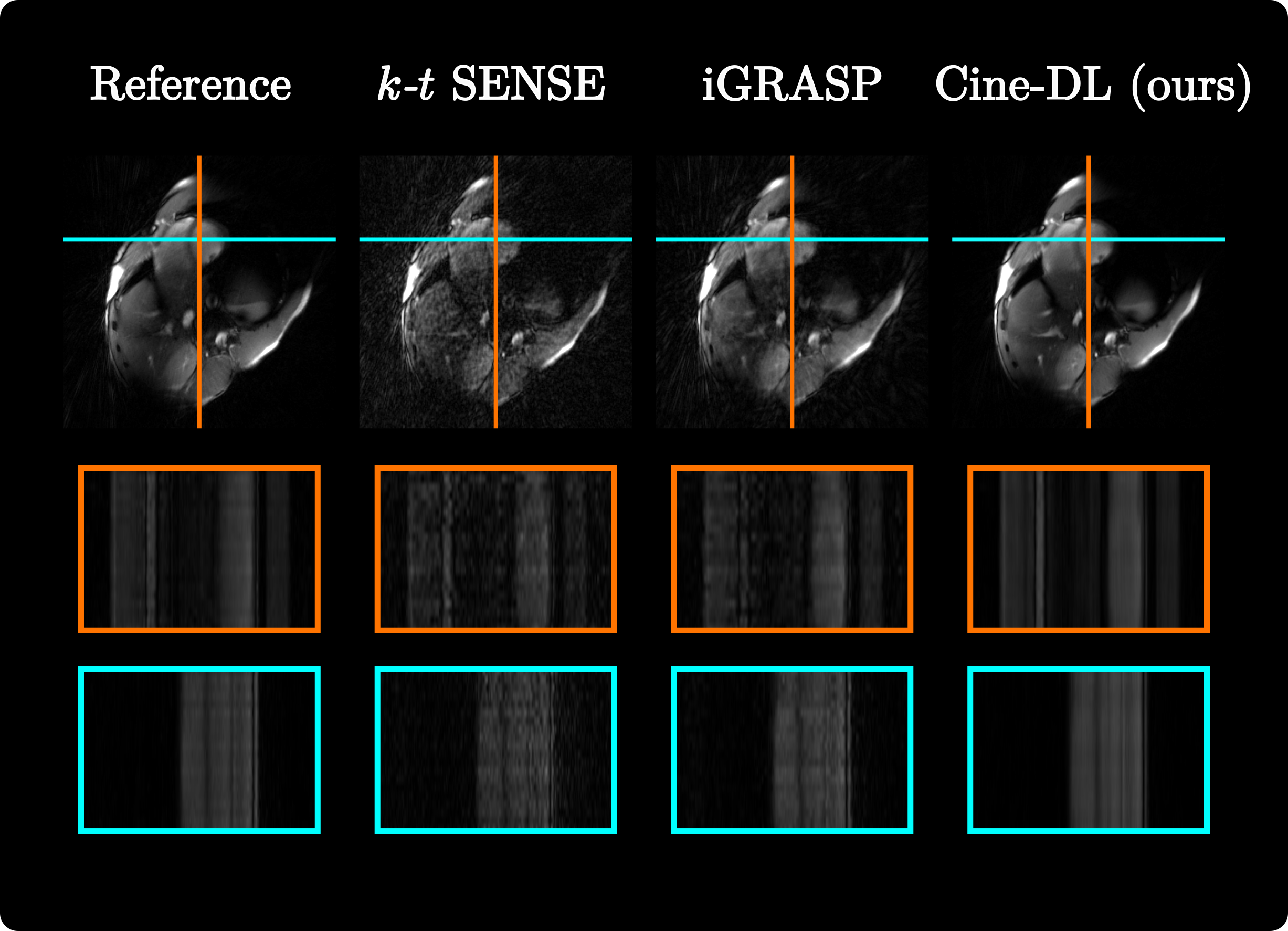}
\caption{\textbf{\textit{x-t} line-profile comparison of free-breathing radial cine MRI reconstructions.} The reference acquisition (80~s/slice) is shown alongside \textit{k-t}~SENSE, iGRASP, and the proposed Cine-DL at R=6x (13~s/slice). \textit{x-t} profiles are generated by stacking signal intensity over time along the vertical (orange) and horizontal (cyan) lines indicated in the top row. The corresponding \textit{x-t} maps are shown in the middle (orange boxes) and bottom (cyan boxes) rows. \textit{k-t}~SENSE exhibits temporal incoherence and streaking/aliasing that obscure motion patterns, while iGRASP reduces artifacts but retains residual noise and temporal blurring. Cine-DL more closely matches the reference, preserving sharper temporal transitions and more coherent depiction of cardiac motion.}
\label{fig:xt_results}
\end{figure}

\textbf{Setup: }We next demonstrate end-to-end reconstruction performance on the volunteer dataset. All training data were first pre-processed using the proposed SOC compression to suppress peripheral interference and reduce streak-driving contributions prior to reconstruction. We used respiratory gating and performed cardiac binning with a bin size of 20 cardiac phases. We trained the reconstruction model using retrospectively undersampled data with acceleration factors up to R=8x to expose the network to a range of aliasing/streaking patterns during learning. At test time, we evaluated generalization under R=\{4x, 6x, 8x\} accelerations. Quantitative reconstruction accuracy was measured using PSNR and SSIM (see Section~\ref{sec:metrics} for definitions). In addition, we performed qualitative visual evaluation using reconstructed images and corresponding \textit{x-t} line profiles to assess spatial detail preservation and temporal fidelity. Here, an \textit{x-t} profile is formed by sampling intensity values along a fixed 1D spatial line in each frame and stacking these samples over time, yielding a 2D visualization of motion trajectories across the cardiac cycle. We compared against two established radial cine reconstruction baselines: \textit{k-t} SENSE and iGRASP.
\\~\\
\textbf{Results: }PSNR and SSIM measurements reported in Table~\ref{table:main} show that the proposed method achieves higher PSNR and SSIM than both \textit{k-t} SENSE and iGRASP across all test cases. Qualitatively, representative results shown in Fig.~\ref{fig:retrospective_results} indicate that \textit{k-t} SENSE exhibits residual aliasing and streaking as well as loss of fine anatomical detail, while iGRASP mitigates some artifacts but can introduce smoothing and residual artifacts. Meanwhile, the proposed approach produces cleaner reconstructions with substantially reduced streaking and fewer artifacts propagating into the myocardium and blood pool, while better preserving myocardial borders and intracardiac structure. The accompanying \textit{x-t} line profiles shown in Fig.~\ref{fig:xt_results} further corrobate these observations, where compared with kt-SENSE and iGRASP, our method yields sharper, more temporally consistent motion trajectories with reduced frame-to-frame fluctuations and less temporal blurring. Overall, these results suggest that the proposed pipeline improves both spatial detail preservation and temporal fidelity.

\subsection{Demonstrations on clinic data}

\begin{figure}[htb]
    \centering
    \includegraphics[width=0.48\textwidth]{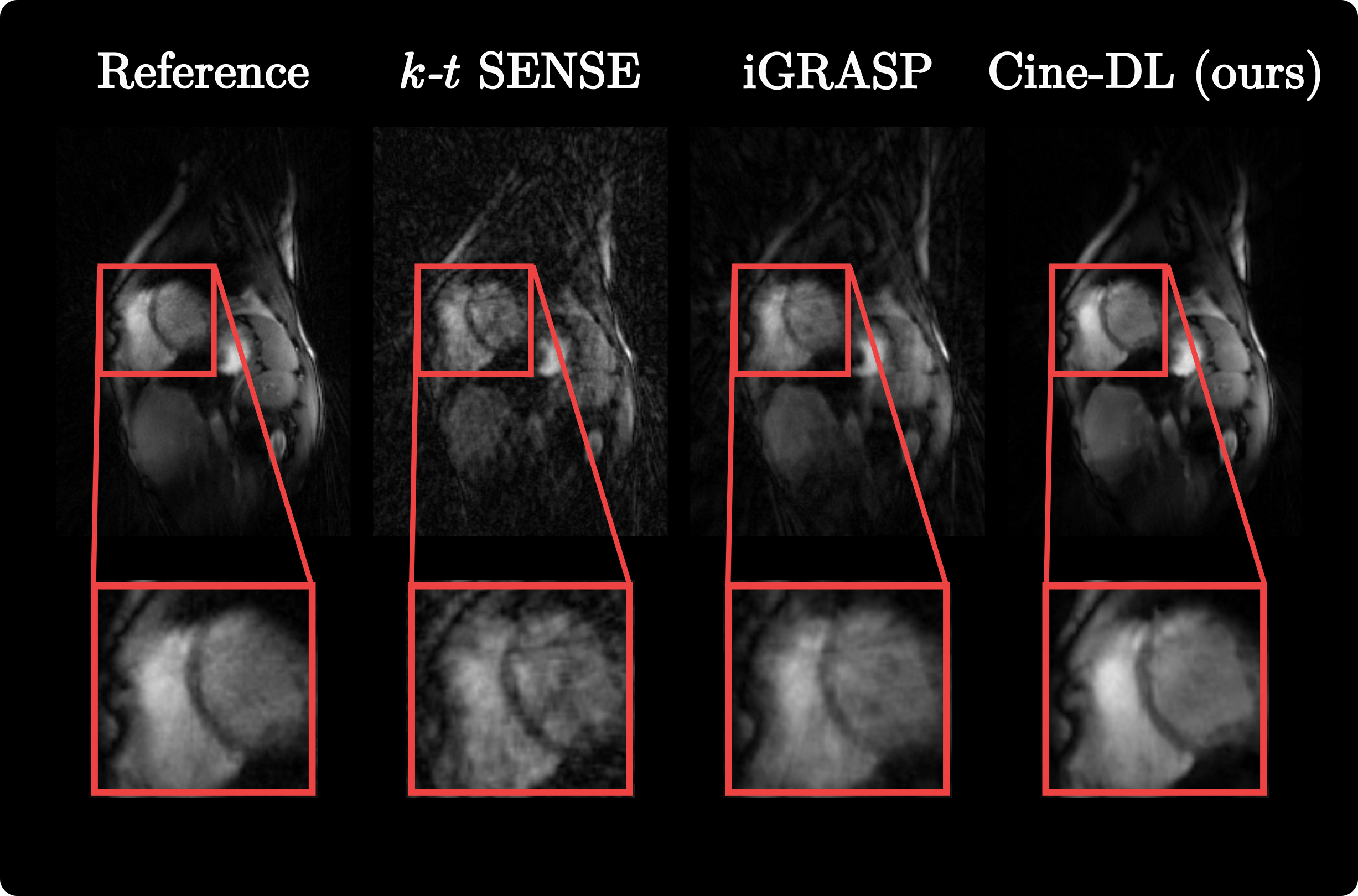}
\caption{\textbf{Clinical retrospective results.} Representative free-breathing radial cine MRI reconstructions from data collected in the clinic. A high-quality reference reconstruction is shown alongside results from retrospectively undersampled data at $R=8\times$. Yellow boxes denote the cardiac region with magnified insets. The proposed Cine-DL produces cleaner images with reduced streaking and fewer artifacts propagating into the myocardium and blood pool, and more closely matches the high-quality reference while preserving sharp myocardial borders and intracardiac structure.}
\label{fig:clinic_retro}
\end{figure}

\textbf{Setup: }We next demonstrate performance on clinical data acquired after deploying our reconstruction pipeline in a local hospital. Data were collected in collaboration with radiologists under routine scanning conditions using an acceleration factor of R=8x. We used the same end-to-end pipeline as in the volunteer experiments, including streak optimized coil compression (SOC) to suppress peripheral interference and reduce streak-driving contributions, followed by respiratory gating and cardiac binning with a bin size of 20 cardiac phases. Reconstructions were produced using the deployed clinical pipeline, and we compare them against standard radial cine baselines used in practice: \textit{k-t} SENSE and iGRASP.
\\~\\
\textbf{Results: }For R=8x acceleration, representative examples (shown alongside reference images) indicate that \textit{k-t} SENSE exhibits residual aliasing/streaking and loss of fine anatomical detail, while iGRASP mitigates some artifacts but can introduce smoothing and residual artifacts. In contrast, the proposed approach produces cleaner reconstructions with substantially reduced streaking and fewer artifacts propagating into the myocardium and blood pool, while better preserving myocardial borders and intracardiac structure, consistent with our volunteer findings. Importantly, the deployed pipeline is feasible in terms of time and memory: SOC reduces channel dimensionality and streak-driving interference prior to reconstruction, and the reconstruction procedure can be executed within practical compute budgets on standard clinical reconstruction hardware (less than 2 minutes per slice). We are currently collecting additional clinical data to enable broader evaluation and more comprehensive radiologist-led assessment.

\label{sec:clinic_results}

\section{Conclusion}
\label{sec:conclusion}

This work introduces Cine-DL, a clinically oriented acquisition-to-reconstruction pipeline for free-breathing cardiac cine MRI. The proposed approach targets the key barriers that have limited routine adoption of highly accelerated radial cine: respiratory motion corruption, streak artifacts, and reconstruction cost. In contrast to conventional breath-hold Cartesian cine which is often infeasible in pediatric, elderly, and dyspneic patients, this framework leverages golden-angle radial sampling for continuous acquisition while explicitly addressing the prominent streaking and computational burden associated with non-Cartesian iterative reconstruction. The proposed design couples targeted k-space preprocessing, Streak Optimized Coil Compression (SOC), and a fast model-based deep reconstruction network, and is evaluated both on volunteer data and through deployment on newly acquired clinical exams.

{
    \small
    \bibliographystyle{ieeenat_fullname}
    \bibliography{main}
}

\end{document}